%% file: main.tex
\documentclass[letterpaper]{article} 
\usepackage{aaai24}  
\usepackage{times}  
\usepackage{helvet}  
\usepackage{courier}  
\usepackage[hyphens]{url}  
\usepackage{graphicx} 
\usepackage[T1]{fontenc}
\usepackage{amsmath}
\usepackage{xurl}
\usepackage{adjustbox} 
\usepackage{booktabs}
\usepackage{array}
\usepackage{natbib}  
\usepackage{caption} 
\usepackage{algorithm}
\usepackage{algorithmic}

\usepackage[usenames,dvipsnames]{xcolor}
\input{macro}

\usepackage{newfloat}
\usepackage{listings}
\DeclareCaptionStyle{ruled}{labelfont=normalfont,labelsep=colon,strut=off} 
\floatstyle{ruled}
\newfloat{listing}{tb}{lst}{}
\floatname{listing}{Listing}
\title{Retrieval Augmentation via User Interest Clustering}
\author {
    Hanjia Lyu\textsuperscript{\rm 1}, Hanqing Zeng\textsuperscript{\rm 2}, Yinglong Xia\textsuperscript{\rm 2}, Ren Chen\textsuperscript{\rm 2}, Jiebo Luo\textsuperscript{\rm 1}\\
}
\affiliations {
    \textsuperscript{\rm 1}University of Rochester,
    \textsuperscript{\rm 2}Meta\\
    hlyu5@ur.rochester.edu, \{zengh,yxia,renchen\}@meta.com, jluo@cs.rochester.edu\\
}

\usepackage{bibentry}

\begin{document}

\maketitle

\input{content/0_abstract}

\input{content/1_introduction}

\input{content/2_related_work}

\input{content/3_problem_statement}

\input{content/4_method}

\input{content/5_experiments}

\input{content/6_conclusion}

\bibliography{aaai24}

\appendix

\end{document}

%% file: macro.tex
\usepackage{url}
\usepackage{xurl}
\usepackage{graphicx}
\usepackage{booktabs}
\usepackage{amsmath}
\usepackage{subcaption}
\usepackage{xcolor}
\usepackage{adjustbox} 
\usepackage{multirow}
\usepackage{tikz}
\usepackage[inline]{enumitem}
\usepackage{xspace}
\usepackage{dsfont}
\usepackage{soul}
\newcommand{\ie}{\textit{i}.\textit{e}., }
\newcommand{\eg}{\textit{e}.\textit{g}., }

\usepackage{titletoc}
\usepackage{bm}

\definecolor{bg_blue}{RGB}{213,227,251}
\definecolor{bg_yellow}{RGB}{250,243,187}
\definecolor{bg_purple}{RGB}{177,167,207}
\definecolor{bg_red}{RGB}{200,169,188}
\definecolor{bg_green}{RGB}{192,213,175}
\definecolor{bg_skin}{RGB}{245,232,210}

\definecolor{red_color}{RGB}{255,0,0}

\definecolor{yellow_color}{RGB}{255,202,47}

\definecolor{purple_color}{RGB}{64,103,139}

\definecolor{dark_red}{RGB}{153, 31, 41}

\definecolor{green_color}{RGB}{130,139,78}
\newcommand{\greentext}[1]{\textcolor{green_color}{#1}}
\definecolor{brown_color}{RGB}{205,90,161}

\definecolor{lg_color}{RGB}{63,147,139}

\definecolor{com_color}{RGB}{0,0,139}

\definecolor{orange_color}{RGB}{255,148,63}

\definecolor{gray_color}{RGB}{169,169,169}

\definecolor{lightgray}{RGB}{220,220,220}

\definecolor{lightgreen}{RGB}{179,207,176}

\definecolor{lightblue}{RGB}{181,209,230}

\newcommand{\model}{\mbox{\sc UIC}\xspace}

%% file: content/0_abstract.tex
\begin{abstract}
    Many existing industrial recommender systems are sensitive to the patterns of user-item engagement. 
    Light users, who interact less frequently, correspond to a data sparsity problem, making it difficult for the system to accurately learn and represent their preferences. On the other hand, heavy users with rich interaction history often demonstrate a variety of niche interests that are hard to be precisely captured under the standard ``user-item'' similarity measurement. 
    Moreover, implementing these systems in an industrial environment necessitates that they are resource-efficient and scalable to process web-scale data under strict latency constraints.
    In this paper, we address these challenges by introducing an intermediate ``interest'' layer between users and items. 
    We propose a novel approach that efficiently constructs user interest and facilitates low computational cost inference by clustering engagement graphs and incorporating user-interest attention. 
    This method enhances the understanding of light users' preferences by linking them with heavy users. By integrating user-interest attention, our approach allows a more personalized similarity metric, adept at capturing the complex dynamics of user-item interactions. The use of interest as an intermediary layer fosters a balance between scalability and expressiveness in the model.
     Evaluations on two public datasets reveal that our method not only achieves improved recommendation performance but also demonstrates enhanced computational efficiency compared to item-level attention models.
    Our approach has also been deployed in multiple products at Meta, facilitating short-form video related recommendation.
\end{abstract}

%% file: content/1_introduction.tex
\section{Introduction}\label{sec:introduction}

In the landscape of digital content and e-commerce, recommender systems have become instrumental in shaping user experiences by curating personalized content and product suggestions~\cite{wang2021cross, su2023beyond,yang2020mixed}. These systems, which focus on analyzing and interpreting user-item interaction data, help improve user engagement and satisfaction. Netflix's recommendation of movies based on viewing history or Amazon's product suggestions tailored to past purchases are quintessential examples of these systems in action~\cite{gomez2015netflix, leino2007case}. However, the efficacy of these systems varies across the user spectrum due to the diversity in user engagement levels.

For users with minimal interaction history, commonly referred to as ``light users,'' their limited interactions contribute to a sparse dataset. This scarcity of data, compounded by the fact that these sparse interactions can also be noisy, poses a significant challenge for recommender systems to accurately identify and learn their preferences. Moreover, the engagement labels for light users represent a minor fraction of the training data.  Such label imbalance often results in generic or irrelevant recommendations. The trained model tends to exhibit popularity bias, prioritizing widely liked items over those that might be more relevant to individual preferences.

``Heavy users'' are distinguished by their extensive interaction histories. While it might seem logical to assume that more data leads to better recommendations, the situation is often more intricate. Traditional recommender systems, which generally depend on simple efficiency-oriented algorithms like cosine similarity-based KNN search, find it challenging to accurately reflect the complex and specific interests of these users. To illustrate, consider the example of two users: $u$, who has a dual interest in dogs and basketball, and $v$, who likes dogs but not basketball. Within a system using a generic, non-personalized distance metric, the optimization process for user $u$ results in the embeddings for `dogs' and `basketball' being brought closer together, indicating their joint relevance to $u$. In contrast, for user $v$, the optimization process adjusts the embeddings to bring `dogs' closer while distancing `basketball', aligning with $v$'s distinct preferences. These scenarios reveal a limitation of conventional recommender systems, emphasizing the necessity for more sophisticated and effective strategies to cater to the wide-ranging needs and tastes of users throughout the engagement spectrum.

The attention mechanism emerges as a promising solution to enhance the modeling capabilities of recommender systems, enabling the processing of user-item interactions with exceptional granularity. This method differentiates itself by dynamically assigning varying weights to items, factoring in the user's specific interaction history. To illustrate, considering the example of heavy users discussed previously, attention mechanisms can adeptly navigate the nuanced preferences of users like $u$ and $v$ by precisely adjusting the importance of items such as `dogs' and `basketball' based on individual engagement patterns. This approach allows for the generation of highly personalized recommendations, addressing the limitations of simpler user-item embedding similarity models which may not capture the full spectrum of user interests.

However, when scaling to large datasets, the computational demands of attention mechanisms become a significant challenge. 
Fast and approximate KNN algorithms (\eg based on KMeans~\cite{kmeans}) is not applicable when we personalize cosine similarity via attention. As a result, serving an attention-based model requires  computing specific weights for every possible user-item combination. While the training phase may leverage sample Softmax for efficiency, the inference stage incurs substantial computational costs. This aspect is particularly pronounced in scenarios involving tens or hundreds of millions of items, where the scale of the dataset intensifies the computational burden.

In this paper, we tackle these challenges and introduce a novel methodology, referred to as \model (\underline{U}ser \underline{I}nterest \underline{C}lustering), aimed at effectively capturing user interests to augment recommender systems. 
Our approach is rigorously evaluated on two public datasets and further validated through its deployment in multiple products at Meta. Note that while the approach evaluated on the datasets, including MovieLens, shares the core methodology with the production version, some modifications were made for its deployment at Meta. 
The cornerstone of our work is twofold:

\textbf{Intermediate-level User Interest Modeling. } We construct item clusters mined via co-engagement graphs to represent user interests. This intermediate-level of interest representation is adept at modeling the intrinsic user preferences embedded within co-engagement graphs. It proves to be particularly effective in catering to the diverse needs of both light and heavy users. 
The robustness and applicability of our approach are shown by the empirical results obtained from public datasets and the deployment in multiple Meta products. 
The implementation notably enhances the recommendation of short-form videos, leading to significant improvements across a range of metrics at substantial scales. 

\textbf{Scalable Inference Scheme. } Departing from traditional item-level attention, our model pivots towards interest-level attention. This strategic shift enables the seamless integration of acceleration techniques such as the KNN algorithm, significantly optimizing the computational efficiency during the inference phase. When compared to conventional item-level attention models, our approach demonstrates a marked reduction in computational complexity, making it a highly scalable solution for real-world applications.

The remainder of this paper is structured as follows: Section~\ref{sec:related_work} offers a review of the literature related to our research.  We define our problem and conduct a more in-depth motivation analysis in Section~\ref{sec:problem_statement}. The details of our methodology, encompassing the user interest modeling, training, and inference stages, are thoroughly discussed in Section~\ref{sec:method}. Section~\ref{sec:exp} presents an extensive empirical evidence substantiating the practical utility of our method. Finally, we discuss and conclude our study in Section~\ref{sec:conclusion}.

%% file: content/2_related_work.tex
\section{Related Work}\label{sec:related_work}

Our work is closely related to user interest modeling, which integrates content analysis and user behaviors to understand and model user interests. 
Prominent approaches in this domain leverage content to discern user interests through semantic analysis. 
\citet{grvcar2005user} emphasize interest-focused browsing history, associating web pages with topics in an ontology to guide users towards their interests. 
\citet{godoy2006modeling} introduce WebDCC, a document clustering algorithm for unsupervised concept learning from web documents to derive user profiles. 
\citet{DBLP:conf/acl/QiWWYY0020} present a hierarchical user interest model for news recommendation that captures diverse interests in news topics. 
Beyond textual content, \citet{you2016picture} explore shared images as indicators of user interests, suggesting an innovative approach to deduce profiles from visual content.

Another research strand focuses on optimizing neural network architectures for accurate interest modeling~\cite{pi2019practice,pi2020search,feng2019deep,li2019multi,jiang2020aspect,shen2024multi}. 
\citet{jiang2020aspect} propose the Multi-scale Time-aware user Interest modeling Network (MTIN), which uses an interest group routing algorithm to generate detailed user interest groups based on interaction sequences, showcasing neural networks' potential in capturing user interest intricacies. 
\citet{pi2019practice} introduce a memory network-based model for long sequential user behavior data, while DIN~\cite{zhou2018deep} utilizes attention to model user interests dynamically concerning a specific item. 
DIEN~\cite{zhou2019deep} extends DIN by introducing an interest-evolving mechanism to capture the dynamic evolution of user interests over time. 
\citet{yang2023debiased} propose the Debiased Contrastive learning paradigm for Recommendation (DCRec), which unifies sequential pattern encoding with global collaborative relation modeling through adaptive conformity-aware augmentation.

Our contribution diverges from these existing methodologies by proposing a framework that not only constructs user interest profiles through the clustering of engagement graphs but also enhances recommendation precision and computational efficiency. Unlike prior works, our approach specifically addresses the challenge of modeling users' preferences by linking them with patterns observed in neighboring users' interactions via co-engagement signals, thereby enriching the model's understanding of user preferences across the spectrum of engagement levels. By fusing the interest-level attention, we refine the similarity metric between users and items, enabling a more nuanced capture of the dynamic interplay in user-item interactions.

%% file: content/3_problem_statement.tex
\section{Engagement-based modeling in Personalized Recommendations}\label{sec:problem_statement}
In this section, we first introduce the problem setup of personalized recommendations. Next, we discuss the limitations of the engagement-based recommender systems. 

\subsection{Problem Setup and Notation}\label{sec:problem_setup}
Let $\mathcal{U}$ be the set of all users, where each user is denoted by $u \in \mathcal{U}$. We denote the set of all items that can be recommended and interacted as $\mathcal{I}$, where each item is represented by $i \in \mathcal{I}$. The historical ratings or implicit feedback $\mathcal{R}$ represents the interactions between users and items. This could be explicit feedback (\eg ratings) or implicit feedback (\eg clicks, views, purchases). The goal of a recommendation system is to predict the preferences of a user for a set of items and recommend the top-N items that the user is most likely to be interested in, based on the user's past behavior. 

\input{fig_table/tab_notations}

\subsection{Motivation Analysis}\label{sec:motivation_analysis}

\subsubsection{Limitations of Engagement-based Recommender Systems}
We conduct an initial investigation with a basic two-tower model~\cite{yang2020mixed} on the MovieLens-1M dataset~\cite{harper2015movielens} to examine the inherent constraints associated with engagement-based recommender systems.

\textbf{Lack of personalization results in inadequate capture of diverse user interest.} 
Despite its straightforward nature, the basic engagement-based recommender system falls short in accurately capturing the nuanced interests of individual users. This limits the system's ability to effectively recognize and adapt to unique or diverse user interests. 
The `dogs' and `basketball' example discussed in the Introduction section illustrates the challenge: the system's current approach lacks the finesse to tailor its recommendations based on the specific likes and dislikes of each user, leading to a one-size-fits-all model that can miss the mark on personal relevance.

\textbf{Vulnerability to popularity bias.}
This above scenario highlights the inherent complexity in developing precise and encompassing embeddings for each user and item, a challenge that is notably prevelant for users with diverse or multifaceted interests. Popularity bias further complicates this challenge, particularly when the interests in question are niche or less common. Such specific interests tend to be underrepresented in the training data, making it exceedingly difficult for the model to recognize and learn these preferences, thereby impacting the quality and personalization of the recommendations offered. Figure~\ref{fig:popularity_analysis} illustrates a long-tail distribution in item popularity, indicating a tendency for models to favor recommending more popular items, despite user interactions frequently involving less popular ones.

\input{fig_table/fig_popularity_analysis}

\input{fig_table/fig_disparities}

\textbf{Disparities in recommendation quality.} Figure~\ref{fig:disparities} shows the quality of the recommendation provided to users with varying levels of engagement. The users with the highest level of interaction (100\%) receive the most accurate recommendations, as shown by the sharp increase in the NDCG@50 score. This disparity suggests that users who interact less frequently with the platform are disadvantaged by a system that inherently favors heavy users, who engage more often. As a result, light users are less likely to be presented with items that truly resonate with their preferences, which can lead to a diminished user experience and potentially lower platform satisfaction and retention rates.

\subsubsection{Potential Improvement}\label{sec:potential_improvement}

These issues can be alleviated through item-level attention for each user. Consider a target user $u$ with embedding $e_{u}$ and a candidate item $i$ with embedding $e_{i}$. A complex attention neural network could be used to generate a weight $\alpha_{ui}$ to adjust the final score which is computed as $\alpha_{ui} \times \text{cosine\_similarity}(e_{u}, e_{i})$. However, this approach still presents two major challenges:

\textbf{High computational cost for industrial-scale data.} Accelerating or approximate KNN algorithms can be applied to reduce the computational cost during the inference stage. It first clusters items into multiple groups and compares the embedding similarity between the target user and the items of a subset of the groups. However, the user-dependent attention re-weighting prevents the use of such techniques. The attention weights vary with each user, making it impossible for clustering items beforehand. Consequently, for each target user, similarity or distance calculations are required for all items in the entire pool, significantly increasing the serving cost, potentially by several orders of magnitude. This attention method has a complexity that scales linearly with the number of items. Additionally, the attention module, often involving a complex neural network, further escalates the computation cost.
 
 \textbf{Optimization challenges.} Implementing a sophisticated attention module complicates the optimization process. Popularity bias and noise in user engagement could significantly impact the convergence quality of the model.

\subsection{Design Choices}
The above analysis informs the elements of an effective model for personalized retrieval. The model should (1) capture and model user interest for personalized recommendations, and (2) facilitate the integration of acceleration algorithms (\eg clustering) to streamline the retrieval process without sacrificing accuracy. In the following section, we introduce our proposed method \model (\underline{U}ser \underline{I}nterest \underline{C}lustering) that embodies these principles.

%% file: fig_table/tab_notations.tex
\begin{table}[t]
    \centering
    \caption{Notations.}
    \label{tab:notations}
    \adjustbox{max width=\linewidth}{
    \begin{tabular}{ll}
    \toprule[1.1pt]
    Notation & Description \\
    \midrule
    $\mathcal{U}$ & the set of users\\
    $\mathcal{I}$ & the set of items\\
    $\mathcal{R}$ & the set of the interactions between users and items\\
    $\mathcal{U}_{i}$ & the set of users that have interacted with item $i$\\
    $\mathcal{I}_{u}$ & the set of items that user $u$ has interacted with \\
    $\hat{\mathcal{R}}_{u}$ & the set of items recommended to user $u$\\
\bottomrule[1.1pt]        
\end{tabular}
}
\end{table}

%% file: fig_table/fig_popularity_analysis.tex
\begin{figure}[t]
    \centering
    \begin{subfigure}{0.49\linewidth}
        \centering
        \includegraphics[width=\textwidth]{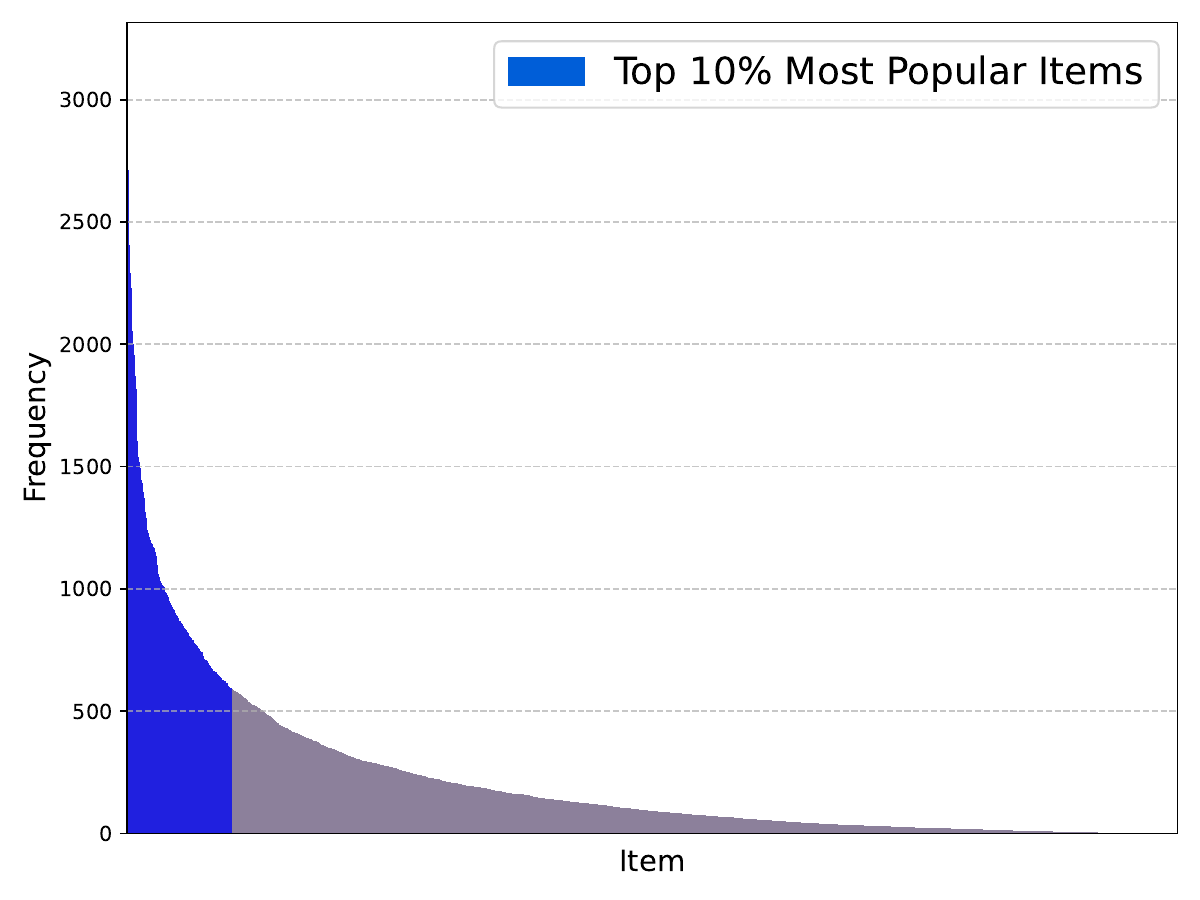}
        \caption{Frequency distribution.}
        \label{fig:popularity_bias}
    \end{subfigure}
    \hfill
    \begin{subfigure}{0.49\linewidth}
        \centering
        \includegraphics[width=\textwidth]{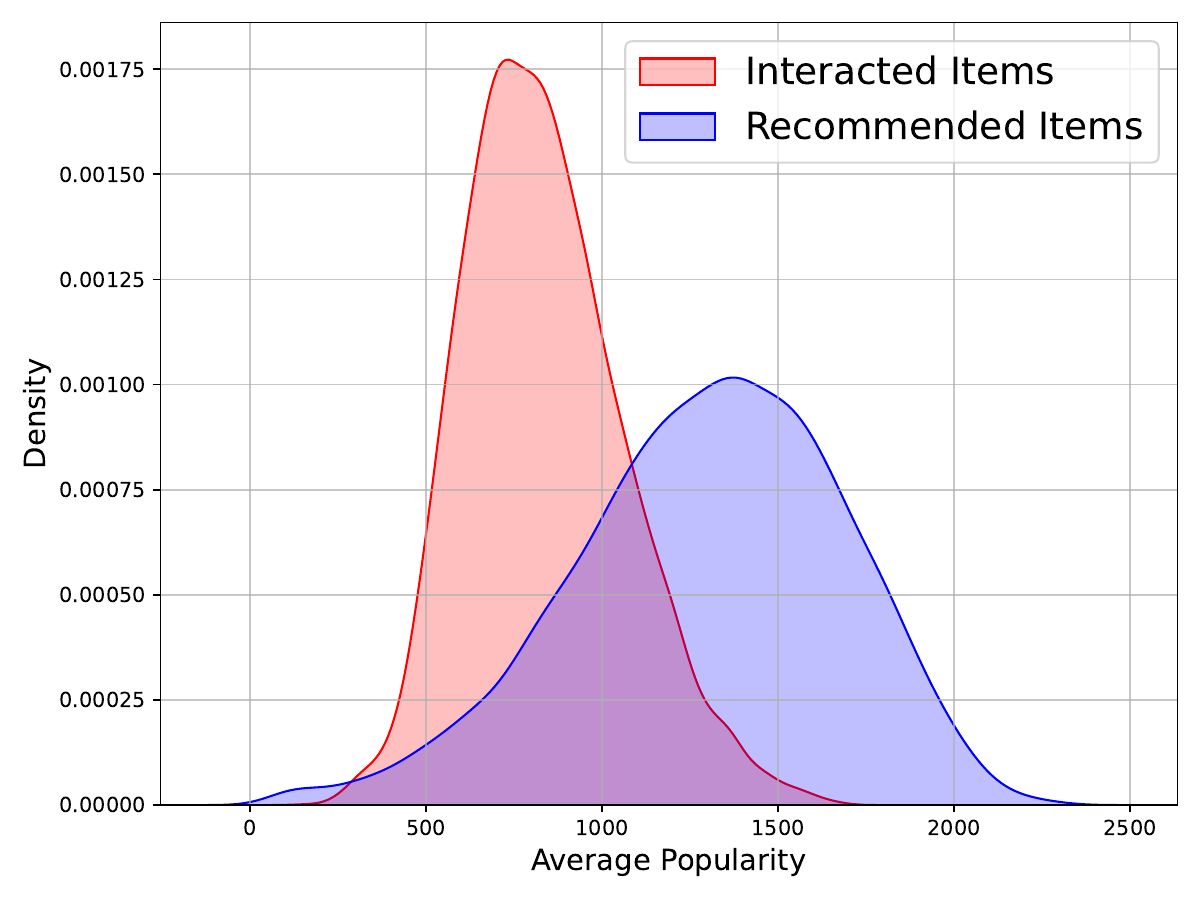}
        \caption{Popularity bias.}
        \label{fig:popularity_diff}
    \end{subfigure}
    \caption{(a) The top 10 most popular items account for almost half of all user-item interactions. (b) The basic two-tower model tends to recommend more popular items, even though users often interact with less popular ones.}
    \label{fig:popularity_analysis}
\end{figure}

%% file: fig_table/fig_disparities.tex
\begin{figure}[ht]
    \centering
    \includegraphics[width=0.75\linewidth]{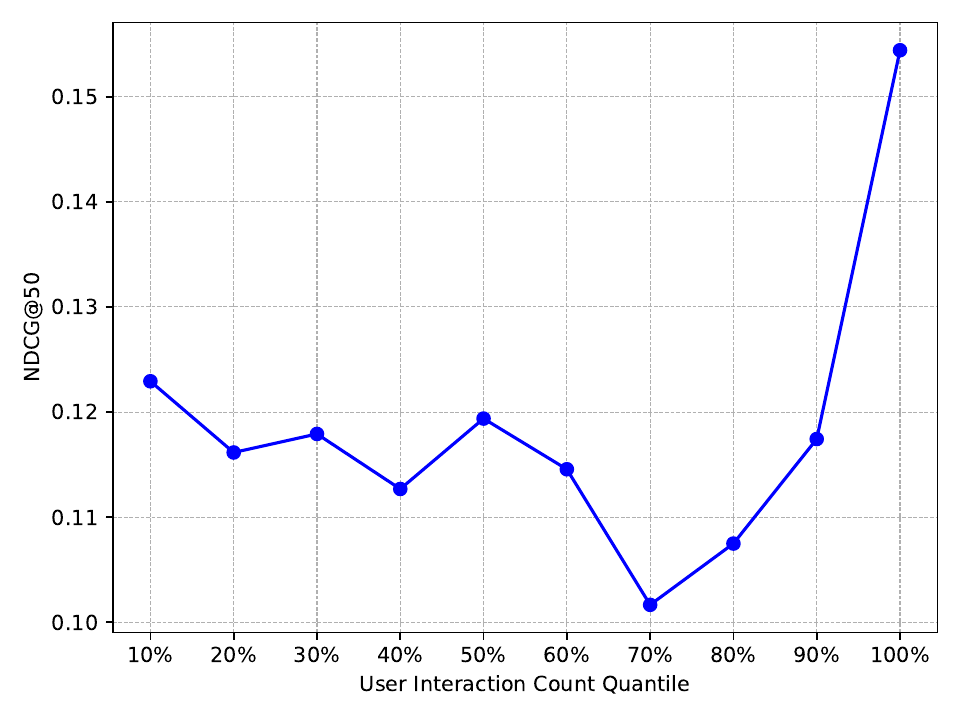}
    \caption{Recommendation quality varies significantly across users with different levels of engagement.}\label{fig:disparities}
\end{figure}

%% file: content/4_method.tex
\section{Methodology}\label{sec:method}
Our approach represents a combination of clustering techniques and attention mechanisms, structured into three distinct stages: interest modeling, training, and inference. During the interest modeling stage, we refine the conventional item-level attention to an intermediate, interest-level attention by clustering co-engagement graphs. This shift simplifies the complexity of ``user-item'' interactions to more manageable ``user-interest'' interactions. The training stage follows the standard practices of engagement-based models. In the inference stage, we implement a KNN method.

\subsection{User Interest Modeling}\label{sec:user_interest_modeling}

\subsubsection{User Interest Clustering}\label{sec:user_interest_clustering}
To reduce the computational burden of item-level attention and enhance the model's resilience to data quality issues, we propose the integration of an intermediate ``interest layer,'' where an ``interest'' represents a cluster of items. 
We consider the user-item bipartite graph $\mathcal{G}=(\mathcal{U}, \mathcal{I}, \mathcal{R})$ where $\mathcal{U}$ and $\mathcal{I}$ denote the sets of user and item nodes, respectively. An edge $e \in \mathcal{R}$ between a user $u \in \mathcal{U}$ and an item $i \in \mathcal{I}$ represents an engagement. Edges are undirected. To cluster items, we first transform $\mathcal{G}$ into an item-item graph $\mathcal{G}_{item \leftrightarrow item}=(\mathcal{I}, \mathcal{E}_{user})$ where edges $\mathcal{E}_{user}$ represent co-engagement. An edge is drawn between item $i$ and item $j$ if there are co-engaged by at least one common user. For simplicity, we do not assign weights to edges. Next, we employ the Louvain communitiy detection algorithm to categorize the items into $K$ different interest clusters: $c_{1}, c_{2}, \ldots, c_{K}$, 
where $c_{i}$ represents the $i$-th interest cluster. The Louvain algorithm is a simple yet effective approach to identifying the community structure within a network.  It operates on a heuristic method that optimizes modularity, a measure that quantifies the strength of division of a network into clusters, thereby facilitating the emergence of item groupings that likely reflect shared user interests~\cite{louvain}. 

Subsequently, we derive a user $u$'s set of interests from the clusters.  Several strategies can be adopted for this purpose.  One method involves identifying the interests with which $u$ has previously interacted or has mostly interacted. Alternatively, we could  implement the Personalized PageRank algorithm (PPR)~\cite{brin1998pagerank}, which, when applied to the user-item bipartite graphs, aid in identifying the items most relevant to a particular user. The user's interests are then deduced from the items that rank as most significant according to the PPR. The items frequently interacted with by others in the same community may also be recommended , even if the target user has not interacted with them directly, thus potentially uncovering latent interests and providing a more holistic profile of user preferences. 
We use the cluster labels of these items to represent user interest, denoted as $\boldsymbol{\eta}_u=[\eta_1, \eta_2, \cdots, \eta_K]$ where $\eta_j$ represents the preference of $u$ for cluster $c_j$. 
Moreover, we integrate the identified interests into the original user embeddings to enrich the representation of user preferences. This fusion can be achieved through concatenation or by using the attention mechanisms.

\subsection{Training Stage}\label{sec:training_stage}
The main focus of this paper is to demonstrate the advantages of modeling user interest within recommender systems. Our proposed framework is adaptable and not confined to specific underlying recommendation module. For the purpose of exemplifying our framework, we select the two-tower model as a representative structure. The two-tower model is widely recognized in recommender systems~\cite{wang2021cross, su2023beyond,yang2020mixed,lyu2024llm}. It is designed to predict the likelihood of a user interacting with an item. It consists of two main neural components, referred to as ``towers,'' each responsible for processing and transforming the features of users and items into dense, lower-dimensional embeddings.
Once the embeddings are generated, the model computes the likelihood of interaction using the dot product or cosine similarity of the embeddings. In our study, we adopt it to integrate and assess the impact of user interest modeling. 

The user feature vector and the item feature vector are passed through a user tower $h_{user}(\cdot)$ and an item tower $h_{item}(\cdot)$, respectively to obtain the user and item embeddings $\bm{e}_{u}$ and $\bm{e}_{i}$.
The feature vectors for users and items can be used to represent a variety of inputs, including user information (\eg demographics) and item metadata (\eg content description). However, given the focus of our study on user interest modeling through interaction patterns, we use only the identifiers for users and items, converting them into input feature vectors via embedding tables. We use MLP as towers. ReLU is used to introduce the non-linearity.

Assuming we obtain $K$ item clusters, and for each user $u$, we obtain user interest $\bm{\eta}_{u} = [\eta_{1}, \eta_{2}, \ldots, \eta_{K}]$ which is a $K$-dimensional vector, where $\eta_{j}$ represents the preference of $u$ for cluster $c_{j}$. This preference, as discussed earlier, is determined through Personalized PageRank (PPR) values.  Next, we fuse the interest representation $\bm{\eta}_{u}$ and  user input feature $\bm{x}_{u}$ as follows:
\begin{equation}
    \bm{e}_{u} = h_{user}(    \bm{W}_{2}(\bm{W}_{1}\bm{\eta}_{u} \oplus \bm{x}_{u})     ),
\end{equation}
where $\bm{W}_{1}$ and $\bm{W}_{2}$ are trainable weight matrices and $\oplus$ is concatenation. An alternative method for integrating user interest involves the computation of an interest-level attention weight, $\alpha_{u, c}$, as a substitute for direct concatenation. The final prediction for the engagement likelihood is computed with dot product: $\hat{y}_{u, i}=\langle \bm{e}_{u}, \bm{e}_{i} \rangle$.

In scenarios where the attention weight $\alpha_{u, c}$ is used for integration, the result of the dot product is further modified by multiplying it with $\alpha_{u, c}$. The user-item interactions are categorized using binary labels \{0,1\}, and the Binary Cross Entropy Loss is selected as the objective function to optimize the model. This methodological approach aims to enhance the predictive accuracy of user-item engagement likelihood by incorporating a nuanced representation of user interests alongside traditional input features.

\subsection{Inference Stage}\label{sec:inference_stage}
At the inference stage, having already established user $u$'s profile of interests, we selectively sample these interests in accordance with their respective weights. Next, we retrieve items by executing KNN searches within the chosen interest clusters.  

\subsection{Computational Complexity}\label{sec:complexity}
Our approach significantly streamlines the serving cost, confining the KNN operations to a limited subset of clusters. 
The user-dependent attention re-weighting, on the other hand, prevents the use of such techniques. The attention weights vary with each user, making it impossible for clustering items beforehand. Consequently, for each target user, similarity or distance calculations are required for all items in the entire pool, significantly increasing the serving cost, potentially by several orders of magnitude. The time complexity of this line of methods scales linearly with the number of items $|\mathcal{I}|$. Assuming the average number of items within each cluster is $\frac{|\mathcal{I}|}{K}$, and we sample $N$ clusters as interest for each user on average, the time complexity is effectively reduced to $O (N \cdot \frac{|\mathcal{I}|}{K})$. This reduction in computational load can be significant if $K$ and $N$ are chosen such that $K \cdot N \ll |\mathcal{I}|$. Note that we also need to cluster items into $K$ clusters beforehand. However, it is an offline operation and does not contribute to the online computational complexity. Furthermore, if the clustering outcomes are stable, re-clustering of items can be infrequent, further optimizing the overall efficiency. Our approach have two major advantages:

\textbf{Comprehensive representation of user interests.} Interest profiles encompass both prominent and tail interests, albeit with varying weights. By incorporating these profiles as input features, the model maintains constant visibility of all interests, tail interests included. 
In contrast, in the absence of such profiles, tail interests (or the items associated with them) seldom surface during training, appearing in only a limited number of batches. This scarcity can introduce noise into the training process and lead the model to overlook these less prominent interests.

\textbf{Mitigation of popularity bias.} The clustering phase inherently offers a mechanism to counteract popularity bias. By constructing interest clusters based on co-engagement graphs, we can integrate debiasing techniques directly into the clustering algorithm. Examples include debiased  PPR and Louvain clustering with constraints on cluster sizes. These methods inherently diminish the influence of overly popular items, promoting a more balanced representation of both dominant and niche interests within the clusters.

%% file: content/5_experiments.tex
\section{Experiments}\label{sec:exp}
\subsection{Experimental Setup}\label{sec:exp_setup}

\subsubsection{Datasets}\label{sec:dataset} We evaluate the effect of different user interest modeling approaches on two public recommendation datasets. Table~\ref{tab:data_stats} shows the dataset statistics. 
Although the two public datasets we use are not as large as those with millions of entries, they are widely used within the research community for evaluating recommendation performance~\cite{li2023health,luo2024collaborative,xi2023towards}. Furthermore, our method has been evaluated on an industrial dataset that is larger by many orders of magnitude (see Section~\ref{sec:ab_test}).

\begin{itemize}[leftmargin=*]
    \item \textbf{MovieLens-1M}~\citep{harper2015movielens} stands as a widely recognized benchmark for evaluating recommender systems.\footnote{License: \url{https://files.grouplens.org/datasets/movielens/ml-1m-README.txt}} This dataset contains 1,000,209 ratings submitted by 6,040 unique users from the MovieLens community. A total of 3,900 movies were rated. Each user has at least 20 ratings. The ratings are converted into 0 or 1 to indicate whether the user has interacted with the item.  
    
    \item \textbf{Recipe}~\citep{majumder2019generating} is another benchmark for evaluating the performance of recommender systems. Sourced from \url{Food.com}, this dataset contains recipe details, reviews and metadata such as ratings, reviews, recipe names, descriptions, ingredients, directions, and so on.\footnote{License: \url{https://www.kaggle.com/datasets/shuyangli94/food-com-recipes-and-user-interactions}} Similar to the MovieLens-1M dataset, we apply filtering criteria. Specifically, we exclude users with fewer than 20 ratings and items with fewer than 10 ratings.
\end{itemize}

\input{fig_table/tab_dataset_stats}

\input{fig_table/tab_main_tab_accuracy}

\subsubsection{Baselines}\label{sec:baseline} We compare our approach against a variety of user interest modeling approaches. To make fair comparison, we use the two-tower model as the same backbone module. We also include a popularity based recommendation method for reference.
\begin{itemize}[leftmargin=*]
    \item \textbf{Most Popular.} This is a simple method where the recommendation is only made by recommending the most popular items.
    \item \textbf{Vanilla.} This is the vanilla two-tower model which is composed of a user tower and an item tower. The final recommendation is made by computing the dot product between the output vectors of the target user and the target items.
    \item \textbf{Vanilla KNN.} It integrates the KNN method discussed in Sections~\ref{sec:inference_stage} and \ref{sec:complexity} to the Vanilla baseline.  KMeans~\cite{kmeans}, Agglomerative Clustering~\cite{mullner2011modern}, and BIRCH~\cite{zhang1996birch} are used for clustering.
    \item \textbf{Item-level Attention.} This applies item-level attention to represent user interest. 
    \item \textbf{Multi-level Attention.} This is one of the state-of-the-art item-level attention mechanism~\cite{zhang2023efficiently}. It applies an Atten-Mixer Network to model user interest based on history interactions.
    \item \textbf{CSR.} Collaborative Sequential Recommendation (CSR)~\cite{luo2024collaborative} leverages both the context within individual user behavior sequences and the collaborative information among the behavior sequences of different users by constructing a local dependency graph.
    \item \textbf{KAR.} Knowledge-Augmented Recommendation (KAR)~\cite{xi2023towards} employs Large Language Models (LLMs) to infer user preferences from their interaction histories with items. Specifically, for each user, a prompt is created to query the LLM to analyze the user's preferences and provide explanations. The generated text serves as a representation of the user's interests. Details on the specific prompt design can be found in the original work.
\end{itemize}

\subsubsection{Evaluation Metrics} We use multiple commonly used metrics to evaluate the performance of our proposed approach and all baselines on Top-K recommendation.
\begin{itemize}[leftmargin=*]
    \item \textbf{Precision.} It measures the relevance of the results recommended by a system. It is defined as the number of relevant items recommended divided by the total number of items recommended.
    \begin{equation}
        \text{Precision@K} = \frac{|\hat{\mathcal{R}}_{u} \cap \mathcal{I}_{u}|}{|\hat{\mathcal{R}}_{u}|}
    \end{equation}
    \item \textbf{Recall.} It is a measure of the system's ability to recommend all relevant items. It is defined as the number of relevant items recommended by the system divided by the total number of relevant items that the user $u$ interacts with.
    \begin{equation}
        \text{Recall@K} = \frac{|\hat{\mathcal{R}}_{u} \cap \mathcal{I}_{u}|}{|\mathcal{I}_{u}|}
    \end{equation}
    \item \textbf{Normalized Discounted Culmulative Gain (NDCG).} NDCG takes the ranks of the correct recommended items into consideration by assigning higher scores to the correct recommended items at top ranks~\cite{he2015trirank, jarvelin2017ir}.
    \begin{equation}
        \text{NDCG@K} = Z_{K} \sum_{k=1}^{K} \frac{I\{\hat{i}_{u, k} \in \mathcal{I}_{u}\}}{\log_2(i+1)}
    \end{equation}
    where $Z_{K}$ is the maximum possible DCG@K for a perfect ranking, $\hat{i}_{u, k}$ denotes the $k$-th recommended item for the user $u$, and $I(\cdot)$ denotes the indicator function.
    \item \textbf{Time.} The inference time across different methods are compared to demonstrate computational efficiency.
\end{itemize}

\subsubsection{Hyper-parameter Settings} 
We randomly set the model parameters, using a Gaussian distribution for initialization. The optimization of the framework is performed using the AdamW algorithm~\cite{loshchilov2017decoupled} with a weight decay value of 0.0005. For all models, we explore learning rates and dropout rates within the ranges of $\{0.0005, 0.001, 0.005\}$ and $\{0.1, 0.3, 0.5\}$. For Kmeans, we set the number of clusters to be \{2\%, 5\%, 10\%\} of the original number of items. To compare it against our approach, we set different resolution values to obtain a similar numbe of clusters. In particular, the resolution is searched over \{1.02, 1.05, 1.1\} and \{1.8, 5.0, 13.0\} for the MovieLens-1M and Recipe datasets, respectively.
We assess performance every 5 epochs and employ an early stopping strategy with a patience setting of 5 to prevent overfitting. The batch size is set at 4096 across all models. We select configurations that yield the highest Recall@50 on the validation dataset for final testing.

\subsubsection{Implementation Details}\label{appendix_sec:implementation details}
Our methods are implemented and experiments are conducted using {\tt PyTorch}.  Each experiment on MovieLens-1M is run on one NVIDIA GeForce RTX 2080 Ti GPU with 11 GB of memory at a time, while the experiments on the Recipe dataset are run on the NVIDIA GeForce RTX 1080 Ti GPU with 11 GB of memory. The clustering algorithms are implemented using the {\tt scikit-learn} package. The Louvain community detection algorithm is implemented using the {\tt community} package.

\subsection{Model Comparison}
Table~\ref{tab:main_accuracy} shows the recommendation accuracy and inference time of {\model} and other baselines, averaged over five different seeds. 

\textbf{Leveraging engagement graphs to construct user interest provides the best balance between recommendation accuracy and computational efficiency.} 
Table~\ref{tab:main_accuracy} shows that {\model} is more accurate, performing best in the Recipe dataset across all accuracy metrics, and achieving the highest precision score in the MovieLens-1M dataset. It ranks second in recall and NDCG for the MovieLens-1M dataset. 
Additionally, it is overall stronger than other clustering algorithms.
More importantly, {\model} is the fastest in terms of computational efficiency. 

Notably, the Item-level Attention method which provides a more fine-grained interest representation yields better performance in recall and NDCG, but not in precision. 
 This discrepancy indicates that while the item-level attention mechanism is adept at generating a larger set of relevant recommendations, it may also introduce items that are less precisely aligned with individual user interests, diluting the precision of recommendations. This method also faces challenges in accurately recommending items within the Recipe dataset, which includes a broader selection of items.

Furthermore, employing a more intricate Item-level Attention model incurs higher computational costs. This increase is due to the necessity of dissecting and processing the subtleties of user-item interactions more thoroughly, which, despite boosting recall and NDCG, escalates the demand on computational resources.

\subsection{Analysis on Light and Heavy Users}\label{sec:light_heavy_analysis}
To investigate whether our proposed model can effectively tackle the disparities in recommendation quality faced by users with varying levels of engagement, we separate users into 10 distinct groups based on their engagement levels. We then employ the Vanilla, Vanilla KNN, and {\model} to generate recommendations. Figure~\ref{fig:light_heavy_comp} illustrates the relative gain in NDCG@50 for {\model} and Vanilla KNN over the Vanilla approach, across users with varying engagement levels.

\input{fig_table/fig_light_heavy_user_comp}

We discover that \textbf{clustering items to represent user interest effectively addresses the challenges associated with serving both light and heavy users.} Overall, both methods prove adept at delivering accurate recommendations for users regardless of their activity level. More importantly, clustering items based on user engagement with the Louvain algorithm proves to be more effective than grouping them by item features, particularly for users with lower levels of interaction.

This distinction suggests that understanding and leveraging the nuances of user engagement allows for a more nuanced and effective recommendation system. For light users, who may not have extensive interaction histories, engagement-based clustering provides a means to infer interests more accurately than relying solely on feature similarity. This method enriches the recommendation pool with items that are likely to resonate with the user's preferences, despite their limited activity. For heavy users, who interact with a wide variety of content, engagement-based clustering helps in identifying nuanced patterns of interest that might be overlooked by content-based methods. This allows for the delivery of more personalized and diverse recommendations that align closely with the user's broad range of interests.

\subsection{Time Comparison}
Figure~\ref{fig:runtime_knn} shows the inference time of the Vanilla method and {\model} on the two datasets. We discover that {\model} consistently surpasses the Vanilla approach in terms of the inference speed across various cluster ratios. The efficiency of {\model} can be attributed to its methodology. Unlike Vanilla approach, which computes rankings for all items within the entire candidate pool—a process that can be computationally intensive—{\model} strategically narrows down the item pool first. It retrieves a targeted subset of items based on the clustered user interests, which reduces the number of comparisons and rankings that the system needs to execute. This streamlined process not only cuts down on computation time but also allows {\model} to maintain or even enhance the quality of recommendations. 

\input{fig_table/fig_knn_cluster_time}

\subsection{Interest Representation}\label{sec:interest_representation}
Figure~\ref{fig:tsne} shows the t-SNE visualization~\cite{van2008visualizing} of the latent vector of the items recommended to a sampled user via \model. Each color in the figure represents a distinct interest cluster. The visualization  demonstrates that items belonging to the same cluster are spatially proximate to one another, indicating the efficacy of {\model}.

\input{fig_table/fig_tsne}

\subsection{Ablation on Number of Clusters}\label{sec:cluster_ablation}

\subsubsection{Number of Clusters during Interest Modeling}
Table~\ref{tab:clustering_ablation_accuracy} shows the recommendation performances of different cluster ratios on the MovieLens-1M dataset during user interest modeling. It reveals that increasing the number of clusters derived from co-engagement graphs enhances the quality of recommendations. 
This underscores the value of directly modeling user interests for improved recommendation accuracy.

\input{fig_table/tab_clustering_ablation_accuracy}

\subsubsection{Number of Clusters during Inference}

Table~\ref{tab:num_clusters_ablation_accuracy} presents the performance outcomes of {\model} when varying the number of clusters selected during the inference stage, with a specific focus on a cluster-to-item ratio of 10\%. Detailed results for experiments conducted with 2\% and 5\% ratios are shown in Tables~\ref{tab:num_clusters_ablation_accuracy_1.02} and \ref{tab:num_clusters_ablation_accuracy_1.05}. The data reveal a trend of improved recommendation accuracy as the number of clusters accessed during inference increases. This aligns with our expectations, as incorporating a larger number of clusters naturally extends the range of items considered, thereby enhancing the system's ability to match user preferences more accurately.

\input{fig_table/tab_num_clusters_ablation}

\input{fig_table/tab_num_clusters_ablation_1.02}

\input{fig_table/tab_num_clusters_ablation_1.05}

\subsection{Optimization Process}\label{sec:opt_process}
The incorporation of user interests may introduce additional complexity to the training optimization process compared to the method that does not integrate any attention mechanisms. 
We perform an experiment comparing the time spent per epoch and the total number of epochs required for both the Vanilla method and our method. 
As shown in Table~\ref{tab:optimization}, our method exhibits faster convergence, requiring fewer epochs to reach optimal performance. Consequently, the overall training time for our method is approximately 2.67\% shorter than that of the Vanilla method, demonstrating its efficiency despite the increased per-epoch duration. 

\input{fig_table/tab_optimization}

\subsection{Maintaining Interest Layers}\label{sec:maintain}

We use the Louvain algorithm to construct user interest clusters based on data with interactions occurring in the earliest 99\%, 98\%, 97\%, 96\%, and 95\% of the time on Movielens-1M, respectively. We then compute the Adjusted Rand Index (ARI)~\cite{santos2009use} of the cluster assignments at two consecutive timestamps. The ARI measures the similarity of two clusterings by considering all pairs of nodes and counting pairs that are assigned to the same or different clusters in the previous and later clusterings. The ARI is adjusted for chance, with a value of 1 indicating identical clusterings and an expected value of 0 for random clusterings. The ARI values for this experiment are 0.84, 0.85, 0.77, 0.75, and 0.80, respectively. These results indicate that the cluster assignment is relatively stable, allowing us to reduce the frequency of constructing and maintaining interest clusters, consequently, decrease the complexity of implementation.

\subsection{Production A/B Test}\label{sec:ab_test}
We have productionized UIC on multiple product surfaces at Meta. 
Considering the huge userbase (billions of monthly active users~\cite{nyt}) of Meta's products (\eg short-form videos), a main challenge in deployment lies in scalability and model execution efficiency. 
Introducing the intermediate ``user interest'' layer to bridge the users and short-form videos is a key step towards fast training and inference -- discovering recent and relevant user interests \textbf{ shrinks the size of the candidate video pool during the retrieval stage.} 
To ensure the freshness of each user's interest profile, we have further implemented an in-house high performance graph processing engine that can efficiently execute the Louvain community detection algorithm~\cite{louvain} in a distributed environment. 
We have developed a variant of the Louvain algorithm to encourage balanced cluster sizes and to enable hierarchical clustering~\cite{DBLP:conf/wsdm/GuDWY20, DBLP:conf/acl/QiWWYY0020} when the users' engagement demonstrate popularity bias.
Note that while the approach evaluated on the datasets, including MovieLens, shares the core methodology with the production version, some modifications were made for its deployment at Meta.

We evaluate our model via short-term (7-day) and long-term online A/B testing. 
We have consistently observed \textbf{ improvement in the quality of the recommended short-form videos}, measured by a diverse set of topline metrics capturing user engagement, content diversity and serving efficiency, providing strong evidence that our proposed model improves personalization by capturing users' niche interests and alleviating popularity bias. 

%% file: fig_table/tab_dataset_stats.tex
\begin{table}[t]
    \centering
    \caption{Statistics of datasets.}
    \label{tab:data_stats}
    \begin{tabular}{cccc}
    \toprule[1.1pt]
    Dataset      & \# Interactions & \# Items & \# Users \\
\midrule
MovieLens-1M & 1,000,209      & 3,706   & 6,040   \\
Recipe       & 319,176        & 21,213   & 4,372  \\
\bottomrule[1.1pt]        
\end{tabular}
\end{table}

%% file: fig_table/tab_main_tab_accuracy.tex
\begin{table*}[t]
    \centering
    \caption{Recommendation accuracy and inference time. The best performance is highlighted in bold and the second best is emphasized with an underline. Relative gains compared to the Vanilla baseline are indicated in \greentext{\textbf{green}}.}
    \label{tab:main_accuracy}
    \vspace{-5pt}
    \adjustbox{max width=\textwidth}{
    \begin{tabular}{lcccccccc}
    \toprule[1.1pt]
              & \multicolumn{4}{c}{\textbf{MovieLens-1M} (\# items = 3,706)}                 & \multicolumn{4}{c}{\textbf{Recipe} (\# items = 21,213)}     \\ \cmidrule{2-9}
              & Precision@50 $\uparrow$   & Recall@50 $\uparrow$      & NDCG@50 $\uparrow$        & Time (s)  $\downarrow$         & Precision@50 $\uparrow$& Recall@50 $\uparrow$& NDCG@50 $\uparrow$& Time (s) $\downarrow$\\ \midrule
Most Popular  &     0.0219    \scriptsize{$\textcolor{gray}{\pm 0.0000}$}        &     0.1006   \scriptsize{$\textcolor{gray}{\pm 0.0000}$}          &     0.0285 \scriptsize{$\textcolor{gray}{\pm 0.0000}$}        & -     &    0.0068 \scriptsize{$\textcolor{gray}{\pm 0.0000}$}             &    0.0562 \scriptsize{$\textcolor{gray}{\pm 0.0000}$}          &   0.0285 \scriptsize{$\textcolor{gray}{\pm 0.0000}$}     &    -  \\
Vanilla            & 0.0485 \scriptsize{$\textcolor{gray}{\pm 0.0002}$} & 0.1884 \scriptsize{$\textcolor{gray}{\pm 0.0017}$} & 0.1184 \scriptsize{$\textcolor{gray}{\pm 0.0008}$} & 2.5402 \scriptsize{$\textcolor{gray}{\pm 0.1053}$} &   0.0075 \scriptsize{$\textcolor{gray}{\pm 0.0001}$}           &   0.0625 \scriptsize{$\textcolor{gray}{\pm 0.0008}$}        &   0.0325 \scriptsize{$\textcolor{gray}{\pm 0.0004}$}      &  4.2389 \scriptsize{$\textcolor{gray}{\pm 0.0387}$}    \\
Vanilla Kmeans        &    0.0486   \scriptsize{$\textcolor{gray}{\pm 0.0002}$}          &    0.1883     \scriptsize{$\textcolor{gray}{\pm 0.0017}$}        &    0.1185    \scriptsize{$\textcolor{gray}{\pm 0.0007}$}         &   1.4860      \scriptsize{$\textcolor{gray}{\pm 0.5819}$}        &    \underline{0.0076} \scriptsize{$\textcolor{gray}{\pm 0.0001}$}          &      \underline{0.0633} \scriptsize{$\textcolor{gray}{\pm 0.0010}$}     &   \underline{0.0329} \scriptsize{$\textcolor{gray}{\pm 0.0004}$}      &  \textbf{1.0150}  \scriptsize{$\textcolor{gray}{\pm 0.0465}$}   \\ 
Vanilla Agglo         & 0.0484 \scriptsize{$\textcolor{gray}{\pm 0.0002}$} & 0.1880 \scriptsize{$\textcolor{gray}{\pm 0.0016}$} & 0.1183 \scriptsize{$\textcolor{gray}{\pm 0.0007}$} & 1.5457 \scriptsize{$\textcolor{gray}{\pm 0.3742}$} &   \underline{0.0076} \scriptsize{$\textcolor{gray}{\pm 0.0001}$}           &   0.0632 \scriptsize{$\textcolor{gray}{\pm 0.0010}$}        &   \underline{0.0329} \scriptsize{$\textcolor{gray}{\pm 0.0005}$}      &  \underline{1.1408} \scriptsize{$\textcolor{gray}{\pm 0.0596}$}    \\
Vanilla BIRCH           & 0.0485 \scriptsize{$\textcolor{gray}{\pm 0.0003}$} & 0.1883 \scriptsize{$\textcolor{gray}{\pm 0.0018}$} & 0.1184 \scriptsize{$\textcolor{gray}{\pm 0.0009}$} & \underline{1.4405} \scriptsize{$\textcolor{gray}{\pm 0.2175}$} &   0.0075 \scriptsize{$\textcolor{gray}{\pm 0.0001}$}           &   0.0630 \scriptsize{$\textcolor{gray}{\pm 0.0010}$}        &   0.0327 \scriptsize{$\textcolor{gray}{\pm 0.0005}$}      &  1.9840 \scriptsize{$\textcolor{gray}{\pm 1.7064}$}    \\
Item-level Attention   &   0.0480   \scriptsize{$\textcolor{gray}{\pm 0.0003}$}            &   \textbf{0.1960} \scriptsize{$\textcolor{gray}{\pm 0.0018}$}              &  \textbf{0.1204} \scriptsize{$\textcolor{gray}{\pm 0.0011}$}         &  3.6857 \scriptsize{$\textcolor{gray}{\pm 0.2391}$}    &  0.0073 \scriptsize{$\textcolor{gray}{\pm 0.0002}$}               &     0.0595 \scriptsize{$\textcolor{gray}{\pm 0.0017}$}         &    0.0302 \scriptsize{$\textcolor{gray}{\pm 0.0012}$}             &  9.2244  \scriptsize{$\textcolor{gray}{\pm 0.1214}$}  \\ 
Multi-level Attention & 0.0486 \scriptsize{$\textcolor{gray}{\pm 0.0002}$} & 0.1884 \scriptsize{$\textcolor{gray}{\pm 0.0003}$} & 0.1186 \scriptsize{$\textcolor{gray}{\pm 0.0004}$} & 4.7153 \scriptsize{$\textcolor{gray}{\pm 0.0936}$} &  \underline{0.0076} \scriptsize{$\textcolor{gray}{\pm 0.0001}$} & 0.0626 \scriptsize{$\textcolor{gray}{\pm 0.0008}$} & 0.0326 \scriptsize{$\textcolor{gray}{\pm 0.0006}$} & 14.6791 \scriptsize{$\textcolor{gray}{\pm 0.7960}$} \\

CSR & 0.0488 \scriptsize{$\textcolor{gray}{\pm 0.0002}$} & 0.1879 \scriptsize{$\textcolor{gray}{\pm 0.0009}$} & 0.1186 \scriptsize{$\textcolor{gray}{\pm 0.0006}$} & 9.5293 \scriptsize{$\textcolor{gray}{\pm 0.5257}$} &  \underline{0.0076} \scriptsize{$\textcolor{gray}{\pm 0.0001}$} & 0.0631 \scriptsize{$\textcolor{gray}{\pm 0.0004}$} & 0.0325 \scriptsize{$\textcolor{gray}{\pm 0.0003}$} & 10.5088 \scriptsize{$\textcolor{gray}{\pm 2.1187}$} \\

KAR & \underline{0.0489} \scriptsize{$\textcolor{gray}{\pm 0.0001}$} & 0.1893 \scriptsize{$\textcolor{gray}{\pm 0.0005}$} & 0.1194 \scriptsize{$\textcolor{gray}{\pm 0.0002}$} & 9.0537 \scriptsize{$\textcolor{gray}{\pm 0.5653}$} &  -  & - & - & - \\

\multirow{2}{*}{\model (Ours)}&     \textbf{0.0491} \scriptsize{$\textcolor{gray}{\pm 0.0003}$}           &    \underline{0.1906} \scriptsize{$\textcolor{gray}{\pm 0.0019}$}          &     \underline{0.1198} \scriptsize{$\textcolor{gray}{\pm 0.0012}$}         &   \textbf{1.3127} \scriptsize{$\textcolor{gray}{\pm 0.0130}$}             &     \textbf{0.0078} \scriptsize{$\textcolor{gray}{\pm 0.0001}$}    &    \textbf{0.0648} \scriptsize{$\textcolor{gray}{\pm 0.0011}$}
 &       \textbf{0.0334} \scriptsize{$\textcolor{gray}{\pm 0.0008}$} &   2.2794 \scriptsize{$\textcolor{gray}{\pm 0.0873}$}\\
 & (\greentext{\textbf{+1.24\%}})&(\greentext{\textbf{+1.17\%}})&(\greentext{\textbf{+1.18\%}})& (\greentext{\textbf{-48.32\%}})&(\greentext{\textbf{+4.00\%}})&(\greentext{\textbf{+3.68\%}})&(\greentext{\textbf{+2.77\%}})&(\greentext{\textbf{-46.23\%}})\\
\bottomrule[1.1pt]
\end{tabular}
}
\end{table*}

%% file: fig_table/fig_light_heavy_user_comp.tex
\begin{figure}[t]
    \centering
    \begin{subfigure}{0.49\linewidth}
        \centering
        \includegraphics[width=\textwidth]{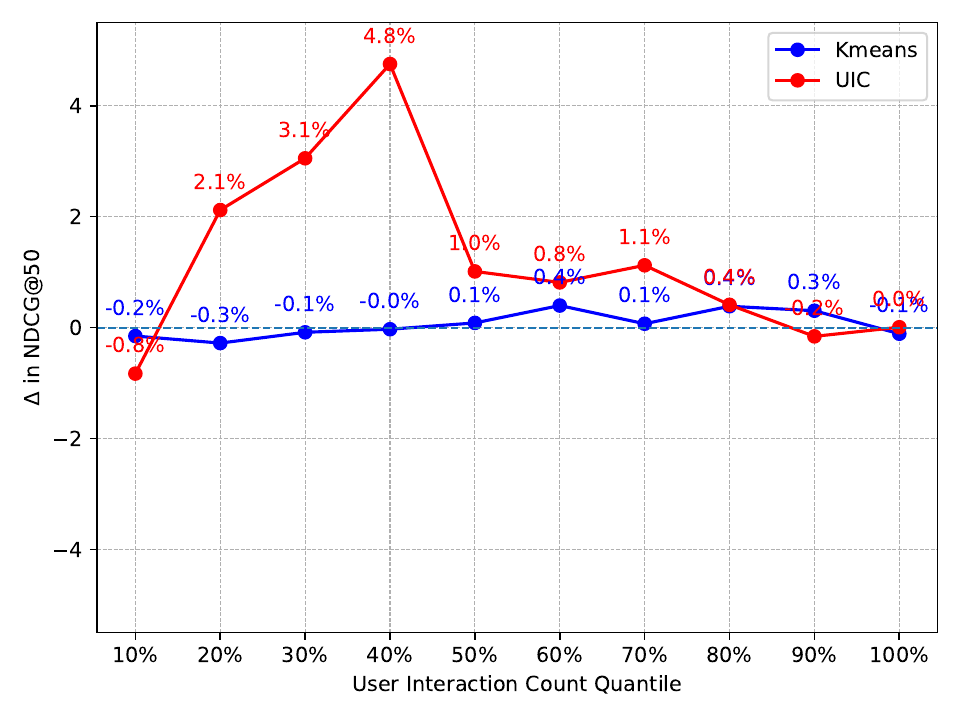}
        \vspace{-15pt}
        \caption{MoveiLens-1M.}
        \label{fig:light_heavy_comp_ml-1m}
    \end{subfigure}
    \hfill
    \begin{subfigure}{0.49\linewidth}
        \centering
        \includegraphics[width=\textwidth]{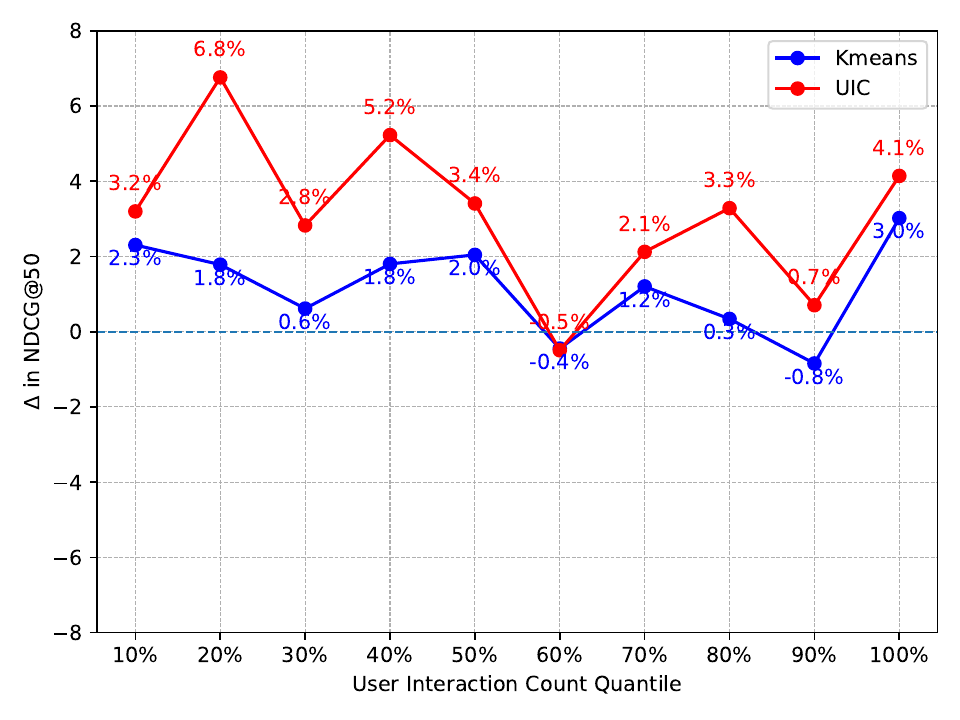}
        \vspace{-15pt}
        \caption{Recipe.}
        \label{fig:light_heavy_comp_recipe}
    \end{subfigure}
    \vspace{-5pt}
    \caption{UIC effectively addresses the challenges associated with serving both light and heavy users. The clustering approach based on engagement signals outperforms the one that relies solely on item features (\ie Kmeans).  The x-axis indicates the percentage of users with lower engagement levels than the referenced user.}
    \label{fig:light_heavy_comp}
\end{figure}

%% file: fig_table/fig_knn_cluster_time.tex
\begin{figure}[t]
    \centering
    \begin{subfigure}{0.49\linewidth}
        \centering
        \includegraphics[width=\textwidth]{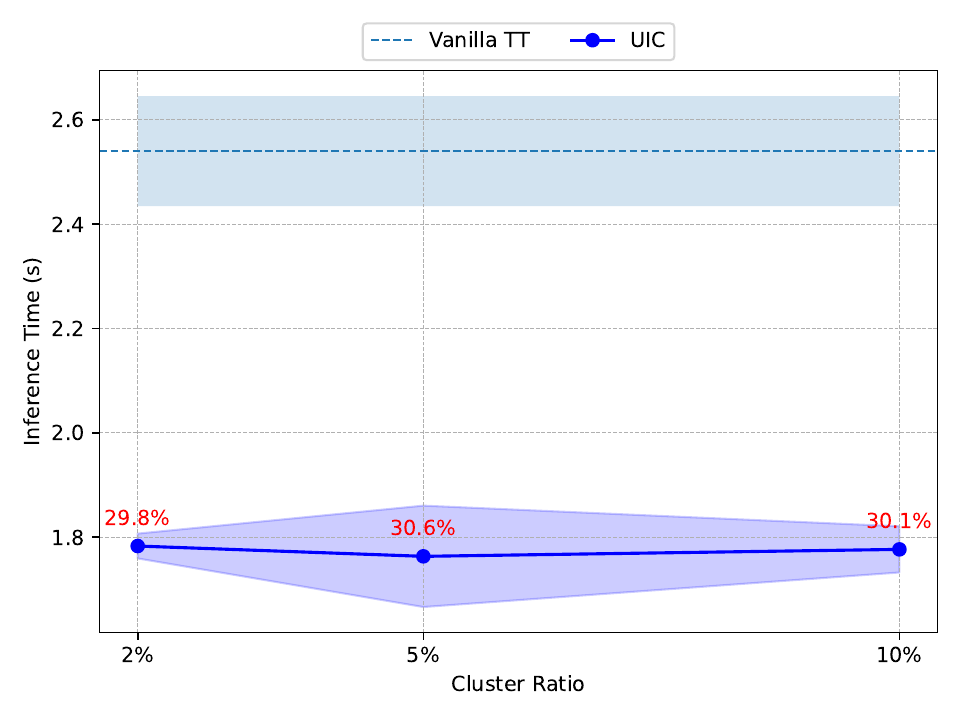}
        \vspace{-15pt}
        \caption{MoveiLens-1M.}
        \label{fig:runtime_knn_ml-1m}
    \end{subfigure}
    \hfill
    \begin{subfigure}{0.49\linewidth}
        \centering
        \includegraphics[width=\textwidth]{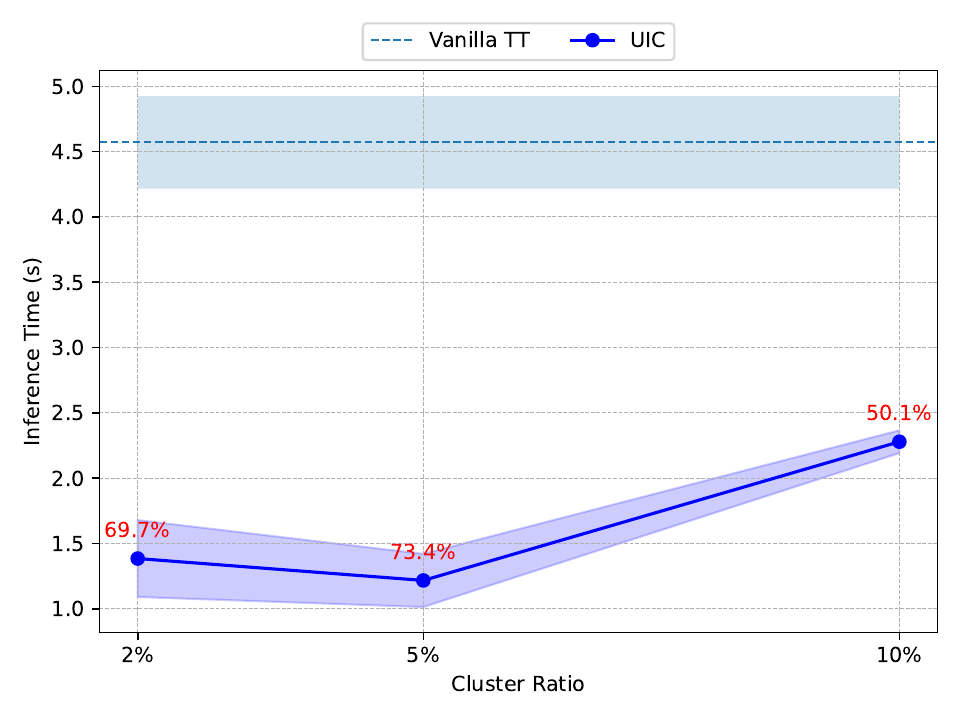}
        \vspace{-15pt}
        \caption{Recipe.}
        \label{fig:runtime_knn_recipe}
    \end{subfigure}
    \vspace{-5pt}
    \caption{Leveraging KNN acceleration techniques reduces inference time significantly.}
    \label{fig:runtime_knn}
\end{figure}

%% file: fig_table/fig_tsne.tex
\begin{figure}[ht]
    \centering
    \includegraphics[width=0.65\linewidth]{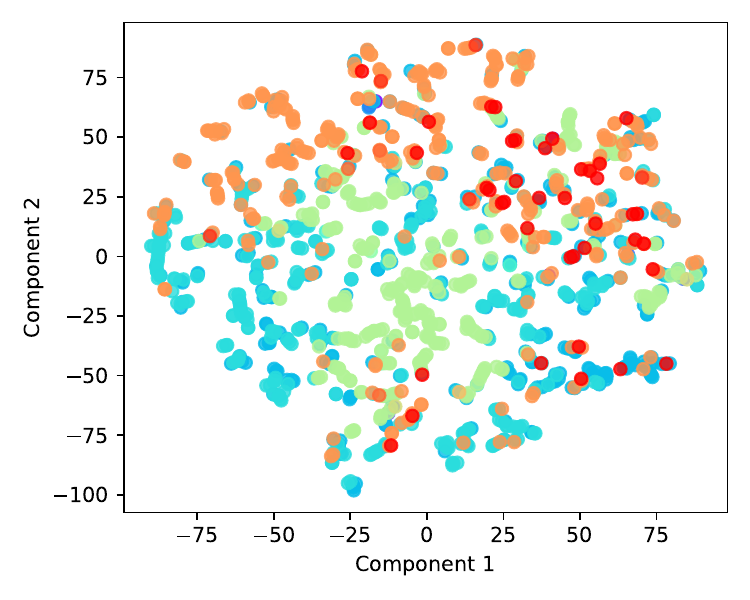}
   \caption{The t-SNE visualization of the item vectors recommended to a sampled user in MovieLens-1M via UIC.}\label{fig:tsne}
\end{figure}

%% file: fig_table/tab_clustering_ablation_accuracy.tex
\begin{table}[t]
    \centering
    \caption{Ablation on number of clusters during user interest modeling.}
    \label{tab:clustering_ablation_accuracy}
    \vspace{-5pt}
    \adjustbox{max width=\linewidth}{
  \begin{tabular}{lccc}
  \toprule[1.1pt]
           & Precision@50 & Recall@50 & NDCG@50 \\ \midrule
Vanilla    &  0.0485 \scriptsize{$\textcolor{gray}{\pm 0.0002}$} & 0.1884 \scriptsize{$\textcolor{gray}{\pm 0.0017}$} & 0.1184 \scriptsize{$\textcolor{gray}{\pm 0.0008}$}        \\
\model (2\%)  &    0.0487     \scriptsize{$\textcolor{gray}{\pm 0.0002}$}     &    0.1895     \scriptsize{$\textcolor{gray}{\pm 0.0006}$}  &   0.1186    \scriptsize{$\textcolor{gray}{\pm 0.0006}$}  \\
\model (5\%)  &      0.0485    \scriptsize{$\textcolor{gray}{\pm 0.0000}$}    &    0.1895     \scriptsize{$\textcolor{gray}{\pm 0.0009}$}  &    0.1194  \scriptsize{$\textcolor{gray}{\pm 0.0004}$}   \\
\model (10\%) &    0.0491  \scriptsize{$\textcolor{gray}{\pm 0.0003}$}        &    0.1906      \scriptsize{$\textcolor{gray}{\pm 0.0019}$} &     0.1198   \scriptsize{$\textcolor{gray}{\pm 0.0012}$}  \\
\bottomrule[1.1pt]
\end{tabular}
}
\end{table}

%% file: fig_table/tab_num_clusters_ablation.tex
\begin{table}[t]
    \centering
    \caption{Ablation on number of clusters during inference (resolution = 1.1). }
    \label{tab:num_clusters_ablation_accuracy}
    \vspace{-5pt}
    \adjustbox{max width=\linewidth}{
  \begin{tabular}{lccc}
  \toprule[1.1pt]
  \# clusters         & Precision@50 & Recall@50 & NDCG@50 \\ \midrule
50    &  0.0481  \scriptsize{$\textcolor{gray}{\pm 0.0002}$} & 0.1861 \scriptsize{$\textcolor{gray}{\pm 0.0011}$} & 0.1181 \scriptsize{$\textcolor{gray}{\pm 0.0008}$}        \\
100  &    0.0485     \scriptsize{$\textcolor{gray}{\pm 0.0002}$}     &    0.1881     \scriptsize{$\textcolor{gray}{\pm 0.0012}$}  &   0.1189    \scriptsize{$\textcolor{gray}{\pm 0.0009}$}  \\
150  &      0.0486    \scriptsize{$\textcolor{gray}{\pm 0.0002}$}    &    0.1889     \scriptsize{$\textcolor{gray}{\pm 0.0014}$}  &    0.1191  \scriptsize{$\textcolor{gray}{\pm 0.0010}$}   \\
200 &    0.0489  \scriptsize{$\textcolor{gray}{\pm 0.0003}$}        &    0.1898      \scriptsize{$\textcolor{gray}{\pm 0.0019}$} &     0.1195   \scriptsize{$\textcolor{gray}{\pm 0.0012}$}  \\
250 &    0.0491  \scriptsize{$\textcolor{gray}{\pm 0.0003}$}        &    0.1906      \scriptsize{$\textcolor{gray}{\pm 0.0019}$} &     0.1198   \scriptsize{$\textcolor{gray}{\pm 0.0012}$}  \\
300 &    0.0490  \scriptsize{$\textcolor{gray}{\pm 0.0003}$}        &    0.1914      \scriptsize{$\textcolor{gray}{\pm 0.0018}$} &     0.1198   \scriptsize{$\textcolor{gray}{\pm 0.0012}$}  \\
\bottomrule[1.1pt]
\end{tabular}
}
\end{table}

%% file: fig_table/tab_num_clusters_ablation_1.02.tex
\begin{table}[ht]
    \centering
    \caption{Ablation on number of clusters during inference (resolution = 1.02). }
    \label{tab:num_clusters_ablation_accuracy_1.02}
    \vspace{-5pt}
    \adjustbox{max width=\linewidth}{
  \begin{tabular}{lccc}
  \toprule[1.1pt]
  \# clusters         & Precision@50 & Recall@50 & NDCG@50 \\ \midrule
10    &  0.0487  \scriptsize{$\textcolor{gray}{\pm 0.0002}$} & 0.1897 \scriptsize{$\textcolor{gray}{\pm 0.0006}$} & 0.1181 \scriptsize{$\textcolor{gray}{\pm 0.0006}$}        \\
20  &    0.0487     \scriptsize{$\textcolor{gray}{\pm 0.0002}$}     &    0.1898     \scriptsize{$\textcolor{gray}{\pm 0.0009}$}  &   0.1187    \scriptsize{$\textcolor{gray}{\pm 0.0007}$}  \\
30  &      0.0487    \scriptsize{$\textcolor{gray}{\pm 0.0002}$}    &    0.1889     \scriptsize{$\textcolor{gray}{\pm 0.0008}$}  &    0.1187  \scriptsize{$\textcolor{gray}{\pm 0.0007}$}   \\
\bottomrule[1.1pt]
\end{tabular}
}
\end{table}

%% file: fig_table/tab_num_clusters_ablation_1.05.tex
\begin{table}[ht]
    \centering
    \caption{Ablation on number of clusters during inference (resolution = 1.05). }
    \label{tab:num_clusters_ablation_accuracy_1.05}
    \vspace{-5pt}
    \adjustbox{max width=\linewidth}{
  \begin{tabular}{lccc}
  \toprule[1.1pt]
  \# clusters         & Precision@50 & Recall@50 & NDCG@50 \\ \midrule
25    &  0.0486  \scriptsize{$\textcolor{gray}{\pm 0.0000}$} & 0.1896 \scriptsize{$\textcolor{gray}{\pm 0.0010}$} & 0.1195 \scriptsize{$\textcolor{gray}{\pm 0.0003}$}        \\
50  &    0.0487     \scriptsize{$\textcolor{gray}{\pm 0.0000}$}     &    0.1903     \scriptsize{$\textcolor{gray}{\pm 0.0009}$}  &   0.1196    \scriptsize{$\textcolor{gray}{\pm 0.0002}$}  \\
75  &      0.0487    \scriptsize{$\textcolor{gray}{\pm 0.0001}$}    &    0.1902     \scriptsize{$\textcolor{gray}{\pm 0.0006}$}  &    0.1195  \scriptsize{$\textcolor{gray}{\pm 0.0001}$}   \\
100 &    0.0487  \scriptsize{$\textcolor{gray}{\pm 0.0001}$}        &    0.1903      \scriptsize{$\textcolor{gray}{\pm 0.0006}$} &     0.1194   \scriptsize{$\textcolor{gray}{\pm 0.0001}$}  \\
125 &    0.0488  \scriptsize{$\textcolor{gray}{\pm 0.0001}$}        &    0.1907      \scriptsize{$\textcolor{gray}{\pm 0.0006}$} &     0.1195   \scriptsize{$\textcolor{gray}{\pm 0.0002}$}  \\
\bottomrule[1.1pt]
\end{tabular}
}
\end{table}

%% file: fig_table/tab_optimization.tex
\begin{table}[t]
    \centering
    \caption{Training time comparison.}
    \label{tab:optimization}
    \vspace{-10pt}
    \adjustbox{max width=\linewidth}{
  \begin{tabular}{lccc}
  \toprule[1.1pt]
           & Total epochs & Total (s) & Time per epoch (s) \\ \midrule
Vanilla    &  250 & 29671.36 &   118.21 \scriptsize{$\textcolor{gray}{\pm 15.69}$}      \\
\model  &    210     &  	28878.69    &   136.87    \scriptsize{$\textcolor{gray}{\pm 17.43}$}  \\
\bottomrule[1.1pt]
\end{tabular}
}
\end{table}

%% file: content/6_conclusion.tex
\section{Discussions and Conclusions}\label{sec:conclusion}

One potential limitation of our approach is the need for additional computing resources during the training stage. However, this requirement may be acceptable given that the model requires less frequent retraining compared to the frequency of inferences, which our method aims to reduce in cost.
Furthermore, clustering can be performed on separate machines, and the Louvain community detection algorithm is computationally efficient as it only considers the graph topology, excluding node and edge features~\cite{louvain}.

In this study, we have introduced {\model}, a method that efficiently constructs user interest profiles and enables cost-effective inference by clustering engagement graphs and leveraging user-interest attention mechanisms. 
Notably, {\model} has been deployed across multiple products at Meta, where it has improved recommendations for short-form videos, achieving substantial gains.